\documentclass[journal,twoside,web]{ieeecolor}
\usepackage{ieee}
\usepackage{cite}
\usepackage{amsmath,amssymb,amsfonts}
\usepackage{graphicx}
\usepackage{textcomp}
\usepackage{algorithm}
\usepackage{algpseudocode}
\usepackage{graphicx} 
\usepackage{multirow}
\usepackage{hhline}

\usepackage[utf8]{inputenc} 
\usepackage[T1]{fontenc}    
\usepackage{hyperref}       
\usepackage{url}            
\usepackage{booktabs}       
\usepackage{amsfonts}       
\usepackage{nicefrac}       
\usepackage{microtype}      
\usepackage{lipsum}
\usepackage{fancyhdr}       
\usepackage{graphicx}       
\usepackage{amsmath}
\usepackage{makecell}
\usepackage{subcaption}
\usepackage{listings}  
\usepackage{xcolor}     
\usepackage{textcomp}
\usepackage{gensymb}
\usepackage{amsmath, amssymb, booktabs, multirow}
\usepackage{cuted}
\usepackage{array}

\newcommand{\bs}{\mathbf}

\let\labelindent\relax
\usepackage{enumitem}

\DeclareMathOperator*{\argmin}{arg\,min}
\DeclareMathOperator*{\argmax}{arg\,max}

\def\BibTeX{{\rm B\kern-.05em{\sc i\kern-.025em b}\kern-.08em
    T\kern-.1667em\lower.7ex\hbox{E}\kern-.125emX}}
\begin{document}
\title{Differentiable Forward and Back-Projector for Rigid Motion Estimation in X-ray Imaging}
\author{Xiao~Jiang, Xin~Wang, Ali~Uneri, Wojciech~B.~Zbijewski, and J.~Webster~Stayman \vspace*{-1.0cm}
\thanks{This work is supported, in part, by NIH grant R01EB030547.}
\thanks{Xiao Jiang, Xin~Wang, Ali~Uneri, Wojciech~B.~Zbijewski, and J.~Webster~Stayman are with the Department of Biomedical Engineering, Johns Hopkins University, Baltimore, MD, 20205 USA. e-mail: \{xjiang43, xwang445, ali.uneri, wzbijewski, web.stayman\}@jhu.edu}}
\maketitle

\begin{abstract}
Objective: In this work, we propose a framework for differentiable forward and back-projector that enables scalable, accurate, and memory-efficient gradient computation for
rigid motion estimation tasks. Methods: Unlike existing approaches that rely on auto-differentiation or that are restricted to specific projector types, our method is based on a general analytical gradient formulation for forward/backprojection in the continuous domain. A key insight is that the gradients of both forward and back-projection can be expressed directly in terms of the forward and back-projection operations themselves, providing a unified gradient computation scheme across different projector types. Leveraging this analytical formulation, we develop a discretized implementation with an acceleration strategy that balances computational speed and memory usage. Results: Simulation studies illustrate the numerical accuracy and computational efficiency of the proposed algorithm. Experiments demonstrates the effectiveness of this approach for multiple X-ray imaging tasks we conducted. In 2D/3D registration, the proposed method achieves $\sim$8× speedup over an existing differentiable forward projector while maintaining comparable accuracy. In motion-compensated analytical reconstruction and cone-beam CT geometry calibration, the proposed method enhances image sharpness and structural fidelity on real phantom data while showing significant efficiency advantages over existing gradient-free and gradient-based solutions. Conclusion: The proposed differentiable projectors enable effective and efficient gradient-based solutions for X-ray imaging tasks requiring rigid motion estimation.
\end{abstract}

\begin{IEEEkeywords}
Differentiable projector, rigid motion, 2D/3D registration, motion compensation, geometry calibration.
\end{IEEEkeywords}

\section{Introduction}

\IEEEPARstart{M}{otion} estimation plays an important role in a wide range of clinical X-ray imaging tasks\cite{kyme2021motion}, such as 2D/3D registration\cite{unberath2021impact,miao2016cnn}, motion-compensated reconstruction\cite{sisniega2017motion,rit2009fly}, and online geometric calibration\cite{ouadah2016self}. In 2D/3D registration, the goal is to align a pre-acquired 3D volume (e.g., a CT scan) with intraoperative 2D X-ray projections\cite{zhang2021long,zhang2022deformable}. This process can be interpreted as estimating the motion between a preoperative reconstruction and the current patient pose. Motion-compensated reconstruction typically involves estimating the object motion first, followed by applying either analytical\cite{capostagno2021deformable} or iterative reconstruction\cite{ouadah2017correction} methods that incorporate the motion model to reduce image artifacts. More advanced techniques seek to jointly perform the motion estimation and image reconstruction to enhance image quality\cite{ouadah2016self,de2025adaptive,de2025solving}. Additionally, motion estimation can be used in geometry calibration tasks, allowing accurate characterization of the gantry trajectory\cite{ma2024fully} if the relative position between source and detector is fixed. 

While deformable motion models\cite{riblett2018data} are often more accurate in capturing complex motion, rigid motion models remain highly relevant in many practical scenarios. For example, rigid models are effective for head imaging\cite{sisniega2017motion}, localizing high-contrast structures such as bones and surgical devices\cite{uneri2017intraoperative}, and providing a local motion model for subsequent deformable registration\cite{huang2024deformable}. Rigid models are also suitable for coarse volume alignment\cite{cui2021planning} and efficient geometry calibration\cite{ma2024fully}, making them an essential component of various clinical workflows.

All the aforementioned tasks involving rigid motion estimation can be formulated as optimization problems, where the objective is to estimate the motion parameters that optimize a chosen loss function or image quality metric. For example, 2D/3D registration typically employs a projection similarity measure to match digitally reconstructed radiographs with real 2D projections\cite{pickering2009new,ghafurian2017computationally,liu20222d,zhang2022deformable}, while motion-compensated reconstruction often uses image sharpness measures to quantify the motion estimation accuracy\cite{sisniega2017motion,wicklein2012image,groen1985comparison,mateos2012comparative,kingston2011reliable,bueno2005fast}. Recent advances in deep learning have also introduced data-driven metrics that provide larger capture ranges and improved robustness\cite{huang2022reference,gao2023fully}.

While extensive prior work has focused on designing effective objective functions, the choice of optimization algorithm is also critical to ensure reliable solutions. Since both registration and reconstruction depend on forward and back-projection operations that are functions of the object motion, computing the gradients of these operations with respect to motion parameters is a critical step for gradient-based optimization. However, these gradient computation can be prohibitively computationally expensive since the large number of path lengths\cite{https://doi.org/10.1118/1.595715} and voxel footprinst\cite{5482021} are typically computed on-the-fly. To circumvent this difficulty, many existing methods rely on gradient-free algorithms\cite{hansen2023cmaevolutionstrategytutorial,10.1093/comjnl/7.2.155,Olsson01021975,1304849} or finite-difference approximations to estimate gradients\cite{https://doi.org/10.1118/1.4941012,https://doi.org/10.1118/1.4873675,10.1007/978-3-319-19387-8_15}. Although these approaches have demonstrated effectiveness, they require multiple forward passes per iteration, thereby compromising computational efficiency\cite{https://doi.org/10.1118/1.4941012}.

\begin{table*}[htbp]
\centering
\caption{Summary of notations used in this work.}
\renewcommand{\arraystretch}{1.1}
\begin{tabular}{lll}
\toprule
\textbf{Groups} & \textbf{Symbols} & \textbf{Descriptions} \\
\midrule
\multirow{8}{*}{Rigid motion}
&$\bs{t} = (t_x,t_y,t_z)^T$ & Translation along the $x,y,z$ axes\\

&$\boldsymbol{\gamma} = (\gamma_x,\gamma_y,\gamma_z)^T$ & Rotation around the $x,y,z$ axes\\

&$\boldsymbol{\theta} = (t_x,t_y,t_z,\gamma_x,\gamma_y,\gamma_z)^T$ & Motion parameters\\

&$\bs{R}(\boldsymbol{\gamma})=    
    \begin{pmatrix}
         R_{11} & R_{12} & R_{13}\\
         R_{21} & R_{22} & R_{23}\\
         R_{31} & R_{32} & R_{33}\\
    \end{pmatrix}$
& Rotation matrix\cite{wiki:Rotation} \\
    
&$\bs{M}(\boldsymbol{\theta}) = [\bs{R}\mid \bs{t}]$ & Rigid motion matrix\cite{wiki:Rigid}\\

&$\bs{H} = \bs{M}^{-1} = [\bs{R}^T \mid -\bs{R}^T \bs{t}]$ & Reverse rigid motion matrix\\

\cmidrule{1-3}

\multirow{3}{*}{Object}
&$\bs{r}' = (x', y', z')^T$ & Object coordinates in object reference frame \\

&$\bs{r} = (x, y, z)^T = \bs{M}\bs{r}'$ & Object coordinates in world reference frame \\

&$\mu = \mu(\bs{r}')$ & Object value at $\bs{r}'$ \\

\cmidrule{1-3}

\multirow{2}{*}{Detector}
&$\bs{d} = (u, v)^T$ & Detector coordinates \\

&$l = l(\bs{d})$ & Projection value at $\bs{d}$ \\

\cmidrule{1-3}

\multirow{2}{*}{System}
&$\bs{P}=
    \begin{pmatrix}
         P_{11} & P_{12} & P_{13} & P_{14}\\
         P_{21} & P_{22} & P_{23} & P_{24}\\
         P_{31} & P_{32} & P_{33} & P_{34}
    \end{pmatrix}$ & Projection matrix\cite{ouadah2016self} \\
\bottomrule
\label{tab:notation}
\end{tabular}
\end{table*}

Several recent strategies have been introduced to achieve a differentiable projector solution that enables fast gradient computation and that can be integrated into gradient-based optimization frameworks. One popular approach is to leverage the auto-differentiation capabilities of modern deep learning frameworks\cite{2017Automatic}. For example, Gopalakrishnan \textit{et al}.~\cite{10.1007/978-3-031-23179-7_1,Gopalakrishnan_2024_CVPR} implements a vectorized ray-driven forward projector in PyTorch and relies on auto-differentiation to compute gradients for 2D/3D registration tasks. Similarly, Huang \textit{et al}.~\cite{10.1117/12.2654361} performs motion-compensated filtered backprojection (FBP) by resampling the backprojected volume according to the current motion estimate and using differentiable interpolation operators to enable effective gradient flow. A related strategy is also applied to 2D/3D registration \cite{gao2023fully} where a differentiable Projective Spatial Transformer (ProST) module is used to model the rigid motion via resampling the pre-scan volume prior to forward projection. In contrast, Thies \textit{et al}.~\cite{10705329} derives an analytical expression of the gradient of the voxel-driven backprojector, demonstrating substantial acceleration in motion-compensated FBP without relying on automatic differentiation.

Although these differentiable solutions have demonstrated fast gradient-based optimization, they still have the following limitations: Auto-differentiation, while simplifying gradient computation, requires retention of the full computational graph and storage of intermediate variables, resulting in high memory consumption\cite{gao2023fully}. This significantly restricts scalability to high-resolution volumes or large numbers of projections, hindering the integration into multi-view registration\cite{ouadah2017correction} or cone-beam CT (CBCT) reconstruction pipelines\cite{de2025adaptive,de2025solving}. Additionally, existing differentiable projectors are typically limited to specific projection models (e.g., ray-driven\cite{Gopalakrishnan_2024_CVPR}, voxel-driven\cite{10705329} approaches). Some applications like CT reconstruction, particularly iterative methods, often rely on more accurate projection models, such as distance-driven\cite{de2004distance} or separable footprint projectors\cite{5482021}. To our knowledge, no differentiable implementations of these sophisticated projectors have been explored. 

We propose a general framework for differentiable forward and back-projector with respect to rigid motion, which can be applied to arbitrary projector algorithms with excellent memory and computational efficiency. Unlike prior work that focuses on differentiability within specific discretized imaging models, we first present an analytical formulation for the gradient of forward and back-projection in the continuous domain. Our key finding is that the gradients of both forward and back-projection operations can be expressed in terms of the forward and back-projection themselves, providing a unified gradient computation scheme across different projector types. Leveraging this analytical formulation, we further describe the discretized implementation along with an acceleration strategy that effectively balances computational speed and memory usage. We conduct simulation studies to validate the gradient accuracy and computational efficiency of the proposed approach, and demonstrate its effectiveness in both simulation and phantom studies for a variety of X-ray imaging tasks.

\section{Method}
\subsection{Continuous Projection Model with Rigid Motion}
\label{sec:continuous}
\subsubsection{Geometry and Notations}
\begin{figure}[h]
    \centering
    \includegraphics[width=\linewidth]{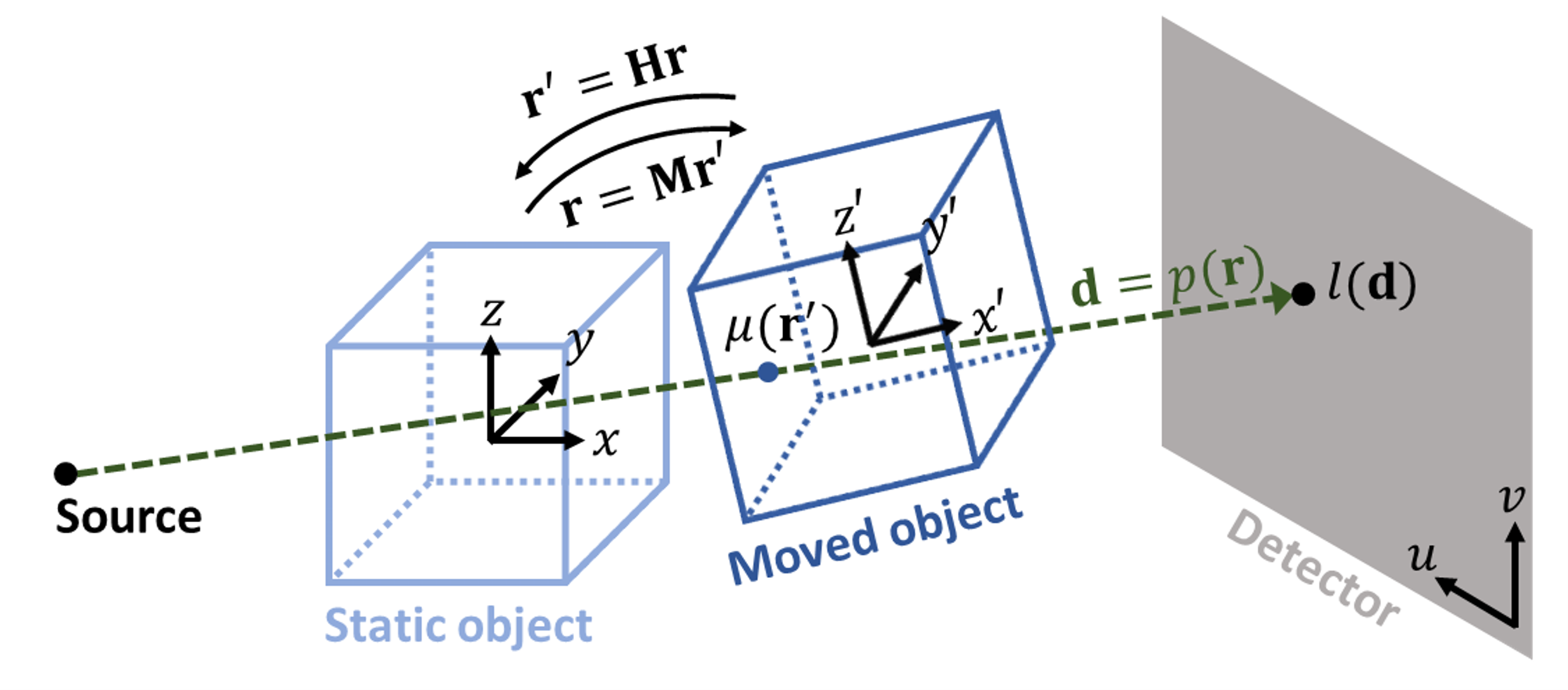}
    \caption{Geometric illustration of X-ray imaging projection model with rigid object motion.}
    \label{fig:geometry}
\end{figure}
We consider a cone-beam projection geometry as shown and parameterized in Fig.\ref{fig:geometry} and TABLE.\ref{tab:notation}. Parameters of rigid object motion are represented by the vector $\boldsymbol{\theta}$, which includes 3 degrees of freedom (DoF) for translation and 3 DoF for rotation. This motion can be represented by a transformation matrix $\bs{M}\in SE(3)$, expressed as a $3\times4$ matrix that concatenates a rotation matrix $\bs{R}\in SO(3)$ and a translation vector $\bs{t}\in \mathbb{R}^3$. We omit the last row of the standard $4\times4$ homogeneous transformation matrix since it is always fixed as $(0,0,0,1)$. The object is described by a 3D distribution function $\mu(\bs{r}')$, where $\bs{r}'$ represents coordinates in the object reference frame, which are related to world coordinates $\bs{r}$ via the transformation $\bs{r}=\bs{Mr}'$, and the inverse mapping is given by $\bs{r}'=\bs{R}^T(\bs{r}-\bs{t})$. We consider a flat-panel detector (FPD), with projection data represented by a 2D distribution function $l(\bs{d})$, where $\bs{d}$ denotes detector coordinates. The projective geometry is described by projection matrix $\bs{P}\in\mathbb{R}^{3\times4}$\cite{ouadah2016self}, which maps a point $\bs{r}$ in 3D space to a point $\bs{d}$ on 2D detector:
\begin{equation}
    \bs{d} = p(\bs{r}):     
    \begin{pmatrix}
         u'\\
         v'\\
         w'
    \end{pmatrix} = \bs{Pr}, \ 
    \bs{d} = 
    \begin{pmatrix} 
        u'/w'\\ 
        v'/w'
    \end{pmatrix}
    \label{eq:projmatrix}
\end{equation}

\subsubsection{Forward and Back-Projector}
A forward projector computes the line integral of a given 3D object, and a backprojector, the adjoint operation of forward projection, places the projection value into the 3D volume along the projection line. Considering rigid object motion, the mathematical expression for forward and back-projection in the continuous domain can be written as: 
\begin{subequations}
\begin{equation}
     \text{Forward:} \ \ \ l(\bs{d}, \boldsymbol{\theta}) = \int \mu(\bs{r}'(\bs{r}, \bs{M}(\boldsymbol{\theta}))) \delta(p(\bs{r}) - \bs{d}) \text{d}\bs{r}, \label{eq:forward}   
\end{equation}
\begin{equation}
    \text{Back:} \ \ \ \mu(\bs{r}', \boldsymbol{\theta}) = \int l(\bs{d}) \delta(p(\bs{r}(\bs{r}', \bs{M}(\boldsymbol{\theta}))) - \bs{d}) \text{d}\bs{d}, \label{eq:back}  
\end{equation}
\end{subequations}
where the the Dirac function mathematically defines the projection line in 3D space, ensuring that only the contributions along the specified line are integrated/backprojected. 

\subsection{Gradient of Projectors with Respect to Motion}
In this section, we derive the gradient of the forward and back-projector with respect to motion parameters, i.e., $\nabla_{\boldsymbol{\theta}}l(\bs{d}, \boldsymbol{\theta})$ and $\nabla_{\boldsymbol{\theta}}\mu(\bs{r}', \boldsymbol{\theta})$, which can be factored as $\nabla_{\boldsymbol{\theta}}l(\bs{d}, \boldsymbol{\theta})=\nabla_{\boldsymbol{\theta}}\bs{M}( \boldsymbol{\theta})\nabla_{\bs{M}}l(\bs{d,M})$ and $\nabla_{\boldsymbol{\theta}}\mu(\bs{r}', \boldsymbol{\theta})=\nabla_{\boldsymbol{\theta}}\bs{M}( \boldsymbol{\theta})\nabla_{\bs{M}}\mu(\bs{r}',\bs{M})$. Since the gradient of the motion matrix with respect to the motion parameters, $\nabla_{\boldsymbol{\theta}}\bs{M}( \boldsymbol{\theta})$, has an explicit expression, we focus on deriving $\nabla_{\bs{M}}l(\bs{d,M})$ and $\nabla_{\bs{M}}\mu(\bs{r}',\bs{M})$. 

\label{sec:gradient}
\subsubsection{Forward Projector}
Based on the forward projection model in Eq.\eqref{eq:forward}, the gradient of forward projection with respect to rigid motion can be derived using the chain rule:
\begin{equation}
\begin{aligned}
    \nabla_{\bs{M}} l(\bs{d}, \bs{M}) &= \nabla_{\bs{M}} \int \mu(\bs{r}'(\bs{r}, \bs{M})) \delta(p(\bs{r}) - \bs{d}) \text{d}\bs{r} \\
        &= \int \nabla_{\bs{M}} \mu(\bs{r}'(\bs{r}, \bs{M})) \delta(p(\bs{r}) - \bs{d}) \text{d}\bs{r} \\
        &= \int \nabla_{\bs{r}'}^T\mu(\bs{r}') \nabla_{\bs{M}} \bs{r}'(\bs{r}, \bs{M}) \delta(p(\bs{r}) - \bs{d}) \text{d}\bs{r}  
\end{aligned}
\label{eq:for_grad}
\end{equation}
The first term in the integrand, $\nabla_{\bs{r}'}^T\mu(\bs{r}')$, represents the spatial gradient of the attenuation map: 
\begin{equation}
    \nabla_{\bs{r}'}^T\mu(\bs{r}')=(\partial_{x'}\mu, \partial_{y'}\mu, \partial_{z'}\mu).
\end{equation}
The Jacobian $\nabla_{\bs{M}} \bs{r}'(\bs{r}, \bs{M})$ can be computed coordinate-wise:
\begin{equation}
    \nabla_{\bs{M}} \bs{r}'(\bs{r}, \bs{M}) = ( \nabla_{\bs{M}} {x}'(\bs{r}, \bs{M}), \nabla_{\bs{M}} {y}'(\bs{r}, \bs{M}), \nabla_{\bs{M}} {z}'(\bs{r}, \bs{M})).
\end{equation}
Given that $\bs{r}'=\bs{R}^T(\bs{r}-\bs{t})$, the gradients for each coordinate are:
\begin{equation}
\begin{aligned}
    \nabla_{\bs{M}} x' &= \begin{pmatrix}
        x - t_x & 0 & 0 & -R_{11} \\
        y - t_y & 0 & 0 & -R_{21} \\
        z - t_z & 0 & 0 & -R_{31}
    \end{pmatrix}, \\
    \nabla_{\bs{M}} y' &= \begin{pmatrix}
        0 & x - t_x & 0 & -R_{12} \\
        0 & y - t_y & 0 & -R_{22} \\
        0 & z - t_z & 0 & -R_{32}
    \end{pmatrix}, \\
    \nabla_{\bs{M}} z' &= \begin{pmatrix}
        0 & 0 & x - t_x & -R_{13} \\
        0 & 0 & y - t_y & -R_{23} \\
        0 & 0 & z - t_z & -R_{33}.
    \end{pmatrix}
\end{aligned}
\end{equation}
Thus, the gradient of the forward projection becomes:
\begin{equation}
    \nabla_{\bs{M}} l(\bs{M},\bs{d})=\int 
    \bs{G}_f(\bs{r})
    \delta(p(\bs{r}) - \bs{d}) \text{d}\bs{r},
\label{eq:final_forward}
\end{equation}
\noindent where the integrand $\bs{G}_f(\bs{r})$ is a $3\times4$ matrix given by:
\begin{equation}
    \bs{G}_f(\bs{r}) = 
    \begin{pmatrix}
        \partial_{x'}\mu & \partial_{y'}\mu & \partial_{z'}\mu
    \end{pmatrix}
    \begin{pmatrix}
        \nabla_{\bs{M}} x' \\
        \nabla_{\bs{M}} y' \\
        \nabla_{\bs{M}} z' \\
    \end{pmatrix}.
\label{eq:for_int}
\end{equation}

\subsubsection{Backprojector}
Based on the backprojection model Eq.\eqref{eq:back}, the gradient of backprojection with respect to rigid motion can be derived using the chain rule:
\begin{equation}
\begin{aligned}
    \nabla&_{\bs{M}} \mu(\bs{r}', {\bs{M}}) = \nabla_{\bs{M}} \int l(\bs{d}) \delta(p(\bs{r}(\bs{r}', {\bs{M}})) - \bs{d}) \text{d}\bs{d} \\
    &= \int l(\bs{d}) \nabla_{\bs{M}} \delta(p(\bs{r}(\bs{r}', {\bs{M}})) - \bs{d}) \text{d}\bs{d} \\
    &= \int l(\bs{d}) \nabla_{\bs{M}} p(\bs{r}(\bs{r}', {\bs{M}})) \delta'(p(\bs{r}(\bs{r}', {\bs{M}})) - \bs{d}) \text{d}\bs{d} \\
\end{aligned}
\label{eq:back_derive}
\end{equation}
Applying properties of the Delta function,
\begin{equation}
    \int f(x)\delta'(x)\text{d}x=-\int f'(x)\delta(x)\text{d}x.
\end{equation}
Eq.\eqref{eq:back_derive} can be further transformed to:
\begin{equation}
\begin{aligned}
    \nabla&_{\bs{M}} \mu(\bs{r}', {\bs{M}}) \\
    &= \int \nabla_{\bs{d}}^T l(\bs{d}) \nabla_{\bs{M}} p(\bs{r}(\bs{r}', {\bs{M}})) \delta(p(\bs{r}(\bs{r}', {\bs{M}})) - \bs{d}) \text{d}\bs{d} \\
    &= \int \nabla_{\bs{d}}^T l(\bs{d}) \nabla_{\bs{r}}^Tp(\bs{r}) \nabla_{\bs{M}}\bs{r} (\bs{r}', {\bs{M}})) \delta(p(\bs{r}(\bs{r}', {\bs{M}})) - \bs{d}) \text{d}\bs{d} \\
\end{aligned}
\label{eq:back_derive}
\end{equation}
The first term in the integrand, $\nabla_{\bs{d}}^T l(\bs{d})$, represents the gradient of the 2D projection: 
\begin{equation}
    \nabla_{\bs{d}}^T l(\bs{d})=(\partial_{u}l, \partial_{v}l).
\end{equation}
The Jacobian of 3D-2D projection mapping function, $\nabla_{\bs{r}}^Tp(\bs{r})$, can be derived from the projection model Eq.\eqref{eq:projmatrix}:
\begin{equation}
\begin{aligned}
    \nabla_{\bs{r}}^Tp(\bs{r}) &=\begin{pmatrix}
                                    \partial_x u & \partial_y u & \partial_z u\\
                                    \partial_x v & \partial_y v & \partial_z v\\
                                \end{pmatrix} \\
    &= \begin{pmatrix}
            \frac{P_{11}w'- P_{31}u'}{w'^2} & \frac{P_{12}w'- P_{32}u'}{w'^2} & \frac{P_{13}w'- P_{33}u'}{w'^2} \\
            \frac{P_{21}w'- P_{31}v'}{w'^2} & \frac{P_{22}w'- P_{32}v'}{w'^2} & \frac{P_{23}w'- P_{33}v'}{w'^2} 
        \end{pmatrix}.
\end{aligned}
\end{equation}
The third term of the integrand of Eq.\eqref{eq:back}, $\nabla_{\bs{M}}\bs{r} (\bs{r}', {\bs{M}}))$, can be computed as the gradient of each coordinate:
\begin{equation}
    \nabla_{\bs{M}} \bs{r}(\bs{r}', \bs{M}) = ( \nabla_{\bs{M}} x(\bs{r}', \bs{M}), \nabla_{\bs{M}} y(\bs{r}', \bs{M}), \nabla_{\bs{M}} z(\bs{r}', \bs{M})).
\end{equation}
Since $\bs{r}=\bs{M}\bs{r}'$, we have:
\begin{equation}
\begin{aligned}
    \nabla_{\bs{M}} x &= \begin{pmatrix}
        x' & y' & z' & 1 \\
        0  & 0  & 0  & 0 \\
        0  & 0  & 0  & 0 \\
    \end{pmatrix}, \\
    \nabla_{\bs{M}} y &= \begin{pmatrix}
        0  & 0  & 0  & 0 \\
        x' & y' & z' & 1 \\
        0  & 0  & 0  & 0 \\
    \end{pmatrix},\\
    \nabla_{\bs{M}} z &= \begin{pmatrix}
        0  & 0  & 0  & 0 \\
        0  & 0  & 0  & 0 \\
        x' & y' & z' & 1 \\
    \end{pmatrix}.
\end{aligned}
\end{equation}
Finally, the gradient of the backprojection can be written as:
\begin{equation}
    \nabla_{\bs{M}} \mu(\bs{r}', {\bs{M}})=\int 
    \bs{G}_b(\bs{d})
    \delta(p(\bs{r}(\bs{r}', M))) - \bs{d}) \text{d}\bs{d},
\label{eq:final_back}
\end{equation}
\noindent where the integrand $\bs{G}_b(\bs{d})$ is a $3\times4$ matrix given by:
\begin{equation}
    \bs{G}_b(\bs{d}) = 
        \begin{pmatrix}
        \partial_u l  & \partial_v l 
    \end{pmatrix}
    \begin{pmatrix}
        \partial_x u & \partial_y u & \partial_z u\\
        \partial_x v & \partial_y v & \partial_z v\\
    \end{pmatrix}
    \begin{pmatrix}
        \nabla_{\bs{M}} x\\
        \nabla_{\bs{M}} y\\
        \nabla_{\bs{M}} z
    \end{pmatrix}.
\label{eq:back_int}
\end{equation}

\subsection{Discretized Implementation}
\label{sec:dicretized}
\subsubsection{Forward Projector}
Although there are various forward projection algorithms\cite{https://doi.org/10.1118/1.595715,bueno2005fast,5482021}, they all follow a unified discretized formulation:
\begin{equation}
    l^i=\sum_ja^{ij}\mu^j, \ \text{or in matrix form:} \ \bs{l=A}\boldsymbol{\mu}
\end{equation}
where the $i, j$ represents the detector pixel index and image voxel index, respectively. The coefficients $a^{ij}$ irepresent voxel-wise weighting factors that account for the voxel footprint and projection path length. Different projection algorithms implement distinct strategies for estimating these weights, resulting in different $a_{ij}$. 

Since the continuous forward model Eq.\eqref{eq:forward} and its gradient Eq.\eqref{eq:final_forward} have exactly the same integration structure, we can conclude that computing the gradient of forward projection is equivalent to forward projecting each element of $\bs{G}_f$ using the same projection algorithm, i.e.,
\begin{equation}
    \nabla_{\bs{M}}l^i=\sum_j a^{ij}\bs{G}^j_f
    \label{eq:dis_forward}
\end{equation}
The volume gradient $\nabla_{\bs{r}'}\mu(\bs{r}')$ in the computation of $\bs{G}_f$ is approximated by a central finite difference. Eq.\eqref{eq:dis_forward} gives the gradient for a single ray. In practice, the projection-domain loss $\mathcal{L}_f$ is usually evaluated over the entire 2D projection, and its gradient with respect to motion is:
\begin{equation}
    \nabla_{\bs{M}} \mathcal{L}_f = \sum_i \nabla_{\bs{M}} l^i \nabla_{l^i} \mathcal{L}_f = \sum_{ij} a^{ij} \bs{G}^j_f \nabla_{l^i} \mathcal{L}_f,
\end{equation}
where $\nabla_{l^i} \mathcal{L}_f$ is the loss gradient with respect to the projection, which can be easily computed given a differentiable loss. The overall gradient computation process can be computed as follows: 

\textbullet\ Compute matrix $\bs{G}_f^j$ for each voxel. 

\textbullet\ Forward project each element of $\bs{G}_f$. 

\textbullet\ Accumulate the projected elements across all projection rays, weighted by the projection loss gradient $\nabla_{l^i} \mathcal{L}_f$.

\subsubsection{Backprojector}
Similar to the forward projection, the unified discretized backprojection model can be written as:
\begin{equation}
    \mu^j=\sum_ia^{ij}l^i, \ \text{or in matrix form:} \ \boldsymbol{\mu}=\bs{A}^T\bs{l}
\end{equation}
Again, the backprojection model Eq.\eqref{eq:back} and its gradient Eq.\eqref{eq:final_back} have exactly the same integration structure, we can conclude that computing the gradient of backprojection is equivalent to backprojecting each element of $\bs{G}_b$ using the same backprojection algorithm. Based on the same principle as in the forward projector, the gradient of the image-domain loss $\mathcal{L}_b$ across the whole 3D volume can be computed as:,
\begin{equation}
    \nabla_{\bs{M}} \mathcal{L}_b=\sum_{ij}a^{ij} \bs{G}^i_b \nabla_{\mu^j} \mathcal{L}_b,
    \label{eq:dis_back}
\end{equation}
where $\nabla_{l^i} \mathcal{L}_b$ is the loss gradient with respect to the voxel value. The projection gradient $\nabla_{\bs{d}}l\bs{(d)}$ in the computation of $\bs{G}_b$ is approximated by a central finite difference. The overall gradient computation process involves: 

\textbullet\ Compute matrix $\bs{G}_b^i$ for each detector pixel.

\textbullet\ Backproject each element of $\bs{G}_b$.

\textbullet\ Accumulate backprojected values across the wole volume, weighted by the voxel-wise loss gradient $\nabla_{\mu^j}\mathcal{L}_b$.

\subsubsection{Efficiency Optimization}
We have developed a PyTorch-compatible GPU-accelerated projector toolbox\cite{jiang2025ctorch} which implements various projector algorithms. Based on this tool, the implementation of Eq.\eqref{eq:dis_forward} and \eqref{eq:dis_back} is straightforward as described in the last section. However, the $\bs{G}_{f/b}$ matrix consumes 12$\times$ memory as the original volume/projection. Alternately, sequential processing can be used to save memory, but will increase runtime by a factor of 12. Because the $\bs{G}_{f/b}$ matrix is only used once per gradient computation, and each of its elements shares the same weighting factors $a^{ij}$ (whose computation is the most time-consuming), we compute $\bs{G}_{f/b}$ on-the-fly and compute $a^{ij}$ once then apply those values to each element of $\bs{G}_{f/b}$. For the on-the-fly computations, atomic operations should be used to avoid the thread conflicts\cite{sanders2010cuda} when summing the gradient from each individual pixel/voxel. Unfortunately, the use of atomic operations compromises computational efficiency. To address this issue, we divide the whole volume/projection into patches, and patch-level shared memory\cite{sanders2010cuda} buffers are allocated to accumulate the gradients locally. The gradients from each pixel/voxel are first written to these local buffers using atomic operations, and then summed into the global gradient matrix. This patch-based strategy allows efficient grid-level parallelization while avoiding excessive thread-level memory allocation, effectively balancing memory usage and computational speed. The projector gradients with respect to motion are implemented using CUDA C and compiled as PyTorch-compatible interfaces, which can be seamlessly integrated into the auto-differentiation framework. 

\subsection{Evaluation}
\label{sec:evaluation}
\subsubsection{Numerical Accuracy}
Since the proposed projectors do not strictly compute the gradient of the discretized projectors, we evaluated their accuracy in a set of validation experiments. We used a $256^3$ static head phantom ($\boldsymbol{\mu}$) from the CQ500 dataset\cite{chilamkurthy2018development} and simulated 8 projections ($\bs{l}$) with projection angles evenly distributed over $0\sim90\degree$. For each projection, we simulated 128 random rigid motions, the translation parameters ($t_x,t_y,t_z$) were sampled from a zero-mean Gaussian distribution with a standard deviation of $20mm$, while the rotation parameters ($\gamma_x,\gamma_y,\gamma_z$) were sampled from a zero-mean Gaussian distribution with a standard deviation of $10\degree$. For each motion realization, we computed the $\mathcal{L}_2$ loss for ray-driven (RD)\cite{https://doi.org/10.1118/1.595715}, distance-driven (DD)\cite{de2004distance}, separable-footprint (SF)\cite{5482021} forward projection and voxel-driven (VD)\cite{trotta2022use}, DD, SF backprojection:
\begin{subequations}
\begin{equation}
    \mathcal{L}_f=\|\bs{A(M(\boldsymbol{\theta}))\boldsymbol{\mu}-l}\|_2^2
    \label{eq:forward_loss}
\end{equation}
\begin{equation}
    \mathcal{L}_b=\|\bs{A}^T\bs{(M(\boldsymbol{\theta})))l-\boldsymbol{\mu}}\|_2^2.
    \label{eq:back_loss}
\end{equation}
\end{subequations}
The gradient with respect to $\boldsymbol{\theta}$ is evaluated using both the proposed algorithm ($\bs{g}_p$) and central finite differences ($\bs{g}_t$), with the latter used as the ground truth. The accuracy of gradient computation was assessed by calculating the cosine similarity between $\bs{g}_p$ and $\bs{g}_t$:
\begin{equation}
    d(\bs{g}_p,\bs{g}_t)=\frac{\bs{g}_p^T\bs{g}_t}{\sqrt{\bs{g}_p^T\bs{g}_p}\sqrt{\bs{g}_t^T\bs{g}_t}}.
\end{equation}

\subsubsection{Computational Cost}
The computational speed and memory consumption was evaluated across four distinct imaging scales, using a volume size $(128n)^3$ and projection size $(192n)^2$, where $n=1,2,3,4$. For the forward projector, the loss \eqref{eq:forward_loss} was evaluated on a single projection. The computational efficiency of the forward projector was then compared with existing methods: DiffDRR\cite{10.1007/978-3-031-23179-7_1} and ProST\cite{gao2023fully}. For the backprojector, the loss \eqref{eq:back_loss} was evaluated on a 256-view projection set, and the computational efficiency was compared with Thies \textit{et al.}\cite{10705329}. It is important to note that all computations included the complete auto-differentiation pipeline, comprising forward loss computation, gradient backpropagation, and gradient descent. All these computations were performed within the PyTorch framework on a personal computer equipped with an AMD Ryzen 9 5950X CPU and a NVIDIA GeForce RTX 4090 GPU.

\subsubsection{Validation on X-ray Imaging Tasks}
\begin{table*}[h!]
    \centering
    \caption{Experiment settings on different X-ray imaging tasks}
    \label{tab:task}
    \renewcommand{\arraystretch}{1.2}
    \begin{tabular}{>{\centering\arraybackslash}m{2.3cm}|>{\centering\arraybackslash}m{3.5cm}>{\centering\arraybackslash}m{3.5cm}>{\centering\arraybackslash}m{3.5cm}}
        \Xhline{1.5pt}
        \textbf{Experiments} & \textbf{2D/3D registration} & \textbf{Motion-compensated analytical reconstruction} & \textbf{Online calibration of noncircular orbits} \\
        \hline
        \textbf{Scan settings} & $10^5$ photons/pixel barebeam, 500 views & 100kVp, 720 views \newline 0.15mAs/view & 100kVp, 360 views \newline 0.15mAs/view \\
        \hline
         & Digital & Real\cite{ma2024fully} & Real \\
        \textbf{Phantom} & \includegraphics[width=0.5\linewidth]{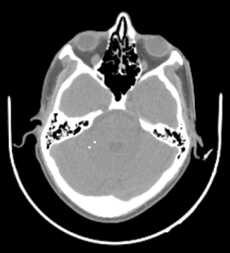} & \includegraphics[width=0.3\linewidth]{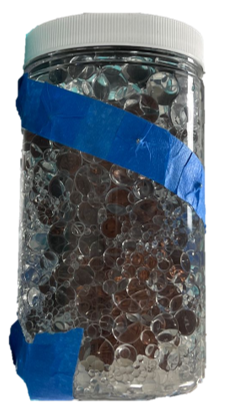} & \includegraphics[width=0.5\linewidth]{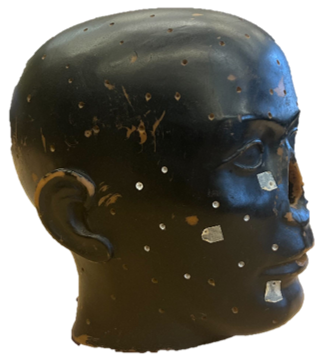}\\
        \hline
        \textbf{Volume size} & 256$\times$256$\times$256 & 256$\times$256$\times$256 & 512$\times$512$\times$512\\
        \textbf{Voxel size} & 1.0$\times$1.0$\times$1.0 mm$^3$ & 0.5$\times$0.5$\times$0.5 mm$^3$ & 0.5$\times$0.5$\times$0.5 mm$^3$ \\
        \hline
        \textbf{Projection size} & 384$\times$384 & 768$\times$768 & 768$\times$768\\
        \textbf{Pixel size} & 0.556$\times$0.556 mm$^2$ & 0.556$\times$0.556 mm$^2$ & 0.556$\times$0.556 mm$^2$ \\
        \hline
        \textbf{Loss function} & \textbf{NCC}\cite{liu20222d} & \textbf{GE}\cite{capostagno2021deformable} & \textbf{NCC + WLS}\cite{ouadah2016self} \\
        \Xhline{1.5pt}
    \end{tabular}
\end{table*}

The performance of the proposed differentiable projector was further evaluated across multiple X-ray imaging tasks: 2D/3D registration, motion-compensated analytical reconstruction, and calibration of a non-circular CBCT geometry for a rigid gantry. Note that the primary objective was to assess the differentiable projector itself, rather than to develop new reconstruction or registration algorithms. Consequently, existing imaging algorithm frameworks were utilized, and the proposed differentiable projectors were plugged in to enable gradient-based solutions. These experiments utilized a ray-driven forward projector and a voxel-driven backprojector, and the configurations are summarized in TABLE.\ref{tab:task} and detailed as follows: 

\textbf{2D/3D registration}: Noisy cone-beam projections (monoenergetic, $10^5$ photons/pixel barebeam, Poisson noise) of a patient model from CQ500 dataset were simulated for 500 views over a circular scan. For each view, random 3 DoF translations ($\sigma_{t_{x/y/z}}=20$ mm) and 3 DoF rotations ($\sigma_{\gamma_{x/y/z}}=20\degree$) were simulated. 2D/3D registration was performed for each view by maximizing the normalized cross-correlation (NCC)\cite{liu20222d} using the proposed differentiable forward projector:
\begin{equation}
   \boldsymbol{\theta}^*=\argmax \ \textbf{NCC}(\bs{A(M(\boldsymbol{\theta}))\boldsymbol{\mu}}_{ref},\bs{l}).
\end{equation}
The motion parameters were initialized to \textbf{0}, and the optimization problem was solved using three algorithms: gradient decent enabled by the proposed projector, gradient descent using DiffDRR, and a derivative-free algorithm using Covariance matrix adaptation evolution strategy (CMA-ES). Gradient-based methods were solved by the Adam optimizer\cite{kingma2014adam}, with step size set to 2$\degree$ for rotations parameters and 10 mm for translations. CMA-ES was initialized with a standard deviation of 10 mm and 5$\degree$ for translations and rotations. Optimization terminated when either the loss change was smaller than $10^{-5}$ or the NCC exceeded 0.999. Registration was considered to be failed
if the optimizer could not achieve a NCC higher than 0.99.

\textbf{Motion-Compensated Analytical Reconstruction}: A spine-and-wire phantom\cite{ma2024fully} was scanned with 720 views on a CBCT test bench\cite{ma2024fully} equipped with a flat-panel detector (Varex PaxScan 4343CB)  a six DoF hexapod robot (Physik Instrumente H-900K Series). The X-ray tube operated in pulse mode with 100kVp tube voltage, 10mA tube current, and 15ms pulse width. During scanning, phantom drift was emulated by gradually shifting the phantom by 1cm along each direction, and rotating by 0.5\degree. The motion-compensated analytical reconstruction\cite{10705329} was formulated as an optimization to maximize the sharpness of the FBP-reconstructed images using gradient entropy (GE) for the sharpness criterion (as in \cite{capostagno2021deformable}). The objective function was:
\begin{equation}
   \boldsymbol{\theta}^*=\argmin \ \textbf{GE}(\bs{A}^T\bs{(M(\boldsymbol{\theta}))}Ramp(\bs{l}))
\end{equation}
where $Ramp$ represents the standard ramp filtering for tomographic reconstruction. The optimization was solved by a gradient-based algorithm using the proposed differentiable projector and the Thies \textit{et al.}\cite{10705329} differentiable projector. Note that the motion parameters were estimated independently for each projection without extra regularization to enforce the smoothness of the motion trajectory. The motion parameters were initialized to \textbf{0}, and updated by the Adam optimizer with 80 iterations. For comparison, CMA‑ES optimization was also evaluated. However, the large number of unknowns ($720\times6$) requires a large population size and leads to ineffective covariance estimation. Following the strategy in \cite{10705329}, a spline‑based motion model was employed: 24 control nodes were evenly distributed over $2\pi$, and motion parameters were obtained by cubic spline interpolation. The whole CMA-ES optimization terminated after $10^4$ function evaluations (527 iterations)\cite{10705329}. 

\textbf{Online Calibration of Noncircular Orbits based on Preoperative CT}: Accurate geometry calibration is essential for cone-beam CT reconstruction, particularly on C-arm scanners where pre-calibrated geometry is often unavailable due to irregular scan trajectories, mechanical jitter, and patient motion \cite{ouadah2016self,ma2024fully}. In this experiment, we acquired 360 projections of an anthropomorphic head phantom (The Phantom Laboratory, Greenwich NY) using a sinusoidal gantry trajectory, emulated by hexapod stage on an X-ray test bench with a maximum tilt angle of 10\degree. The calibration framework with 6 DoF described in Ref.\cite{ouadah2016self} was employed for calibration and reconstruction. Specifically, each projection was registered to a preoperative volume $\boldsymbol{\mu}_{ref}$ using the differentiable forward projector and NCC loss:
\begin{equation}
   \boldsymbol{\theta}^*=\argmin \ \textbf{NCC}(\bs{A(M(\boldsymbol{\theta}))}\boldsymbol{\mu}_{ref},\bs{l}),
\end{equation}
where the motion parameters $\boldsymbol{\theta}$ were initialized with zero. The preoperative volume $\boldsymbol{\mu}_{ref}$ was generated via FBP reconstruction from 1440 projections acquired with a circular scan and no head motion. A Weighted Least Square (WLS) framework was then used for iterative head reconstruction. This step incorporated the estimated motion model from the 2D/3D registration:
\begin{equation}
   \boldsymbol{\mu}^*=\argmin \ \|\bs{A(M(\boldsymbol{\theta}^*))}\boldsymbol{\mu}-\bs{l}\|_\textbf{W}^2,
\end{equation}
where $\bs{W}$ is a statistical weighting determined as the inverse of the sinogram noise variance. NCC and Structural Similarity Index (SSIM) between the reconstruction and preoperative volume were used to quantify the reconstruction accuracy. 

\section{Results}

\subsection{Numerical Accuracy}
\begin{figure}[h]
    \centering
    \includegraphics[width=1.02\linewidth]{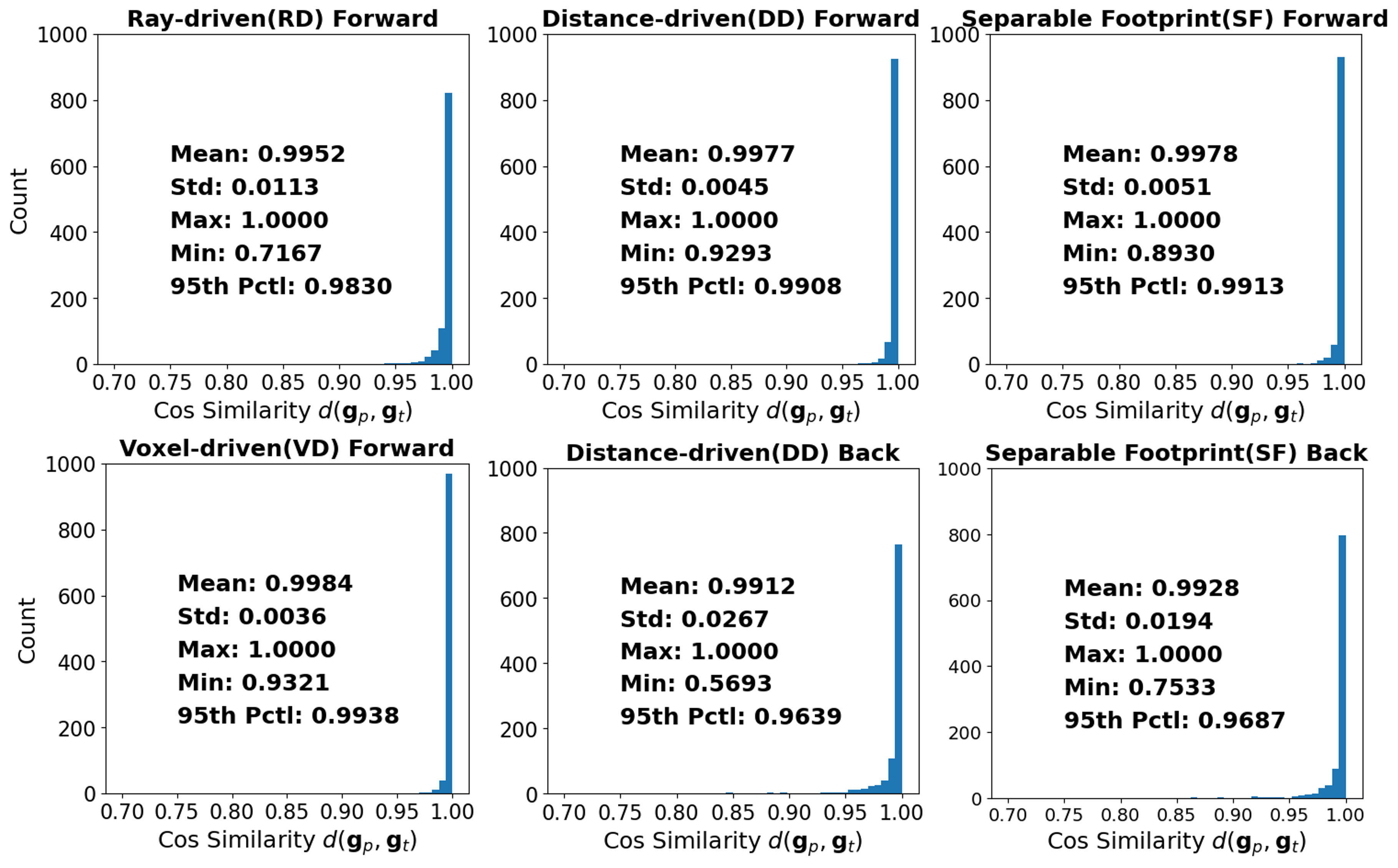}
    \caption{Geometric illustration of X-ray imaging projection model with rigid object motion.}
    \label{fig:accuracy}
\end{figure}
Fig.\ref{fig:accuracy} shows the cosine similarity between the gradient computed using the proposed algorithm ($\bs{g}_p$) and a reference gradient computed using a central finite difference ($\bs{g}_t$) for 1024 random motion realizations. The forward projector achieves mean cosine similarity values of 0.9952, 0.9977, 0.9978 for VD, DD, and SF, respectively, indicating strong agreement with the ground truth. The distributions of cos similarity are heavily skewed towards 1, with high values at the 95th percentile for each projector, suggesting that the proposed forward projectors generally achieve accurate gradient computation. For the backprojection, the VD backprojector achieves excellent accuracy, with a mean cosine similarity of 0.9984 and a 95th percentile of 0.9938. The DD and SF backprojector methods also demonstrate high accuracy, though their distributions are slightly wider and present lower minimum values compared to the VD method.

\subsection{Computational Cost}
\begin{table}[h!]
    \centering
    \caption{Computational cost of differentiable forward projection (1 projection, 100 iterations) with different data sizes (first row: volume size/projection size).}
    \label{tab:forward}
    \renewcommand{\arraystretch}{1.5}
    \begin{tabular}{>{\centering\arraybackslash}m{2.0cm}|>{\centering\arraybackslash}m{1.2cm}>{\centering\arraybackslash}m{1.2cm}>{\centering\arraybackslash}m{1.2cm}>{\centering\arraybackslash}m{1.2cm}}
        \Xhline{1.5pt}
                          & \textbf{128$^3$/192$^2$} & \textbf{256$^3$/384$^2$} & \textbf{384$^3$/572$^2$} & \textbf{512$^3$/768$^2$} \\
        \hline
        \textbf{ProST}\cite{gao2023fully}    & 4.66GB 1.64s & Out of memory & Out of memory & Out of memory \\
        \hline
        \textbf{DiffDRR}\cite{10.1007/978-3-031-23179-7_1}  & 1.21GB 0.79s & 5.74GB 2.55s & 18.07GB 7.69s & Out of memory \\
        \hline
        \textbf{Proposed (RD)} & 0.53GB 1.25s & 0.77GB 2.44s & 1.37GB 6.24s & 2.53GB 12.70s \\
        \hline
        \textbf{Proposed (DD)} & 0.53GB 1.27s & 0.76GB 3.84s & 1.66GB 11.02s & 2.65GB 23.58s \\
        \hline
        \textbf{Proposed (SF)} & 0.58GB 1.41s & 0.80GB 5.03s & 1.68GB 14.49s & 2.70GB 32.53s \\
        \Xhline{1.5pt}
    \end{tabular}
\end{table}

\begin{table}[h!]
    \centering
    \caption{Computational cost of differentiable backprojector (256 views, 1 iteration) with different data sizes (first row: volume size/projection size).}
    \label{tab:back}
    \renewcommand{\arraystretch}{1.5}
    \begin{tabular}{>{\centering\arraybackslash}m{2.0cm}|>{\centering\arraybackslash}m{1.2cm}>{\centering\arraybackslash}m{1.2cm}>{\centering\arraybackslash}m{1.2cm}>{\centering\arraybackslash}m{1.2cm}}
        \Xhline{1.5pt}
                          & \textbf{128$^3$/192$^2$} & \textbf{256$^3$/384$^2$} & \textbf{384$^3$/572$^2$} & \textbf{512$^3$/768$^2$} \\
        \hline
        \textbf{Thies \textit{et al.}}\cite{10705329} & 0.73GB 6.04s & 1.44GB 27.10s & 3.02GB 90.62s & 5.42GB 214.26s \\
        \hline
        \textbf{Proposed (VD)} & 0.57GB 0.17s & 0.95GB 1.21s & 1.83GB 3.99s & 3.50GB 9.50s \\
        \hline
        \textbf{Proposed (DD)} & 0.56GB 0.65s & 0.93GB 4.99s & 1.84GB 16.72s & 3.52GB 39.52s \\
        \hline
        \textbf{Proposed (SF)} & 0.62GB 1.12s & 0.91GB 8.45s & 1.84GB 28.55s & 3.71GB 69.72s \\
        \Xhline{1.5pt}
    \end{tabular}
\end{table}


The computational cost of differentiable forward and back-projection is summarized in TABLE.\ref{tab:forward} and \ref{tab:back}, respectively. For forward projection, ProST and DiffDRR, which both utilize a RD model, quickly exhaust memory as data size increases. In contrast, the proposed RD projector, which benefits from explicit analytical gradient expression instead of auto-differentiation, achieves significant memory reductions over DiffDRR: $56.19\%,86.85\%,92.54\%$ for $128^3/192^2, 256^3/384^2, 384^3/572^2$ data sizes, while maintaining comparable computational speed. Even with a large data size ($512^3/768^2$), the proposed RD projector requires only 2.53GB of memory, highlighting its superior scalability for high-resolution imaging. The proposed DD and SF projectors maintain similar memory usage as the RD projector, with longer computational times due to more sophisticated voxel footprint processing. For backprojection, the proposed VD projector shows substantial efficiency advantages over Thies \textit{et al.}, which also employs a VD model. This significant time difference can be attributed to our acceleration strategy, which utilizes a patch-based gradient buffer to avoid direct accumulation of voxel-wise gradients to global memory. Similar to forward projection, the proposed DD and SF backprojectors exhibit memory consumption comparable to the VD projector but operate with increased computation time.

\subsection{Validation on X-ray Imaging Tasks}
\subsubsection{2D/3D Registration}
\begin{figure}[h!]
    \centering
    \includegraphics[width=\linewidth]{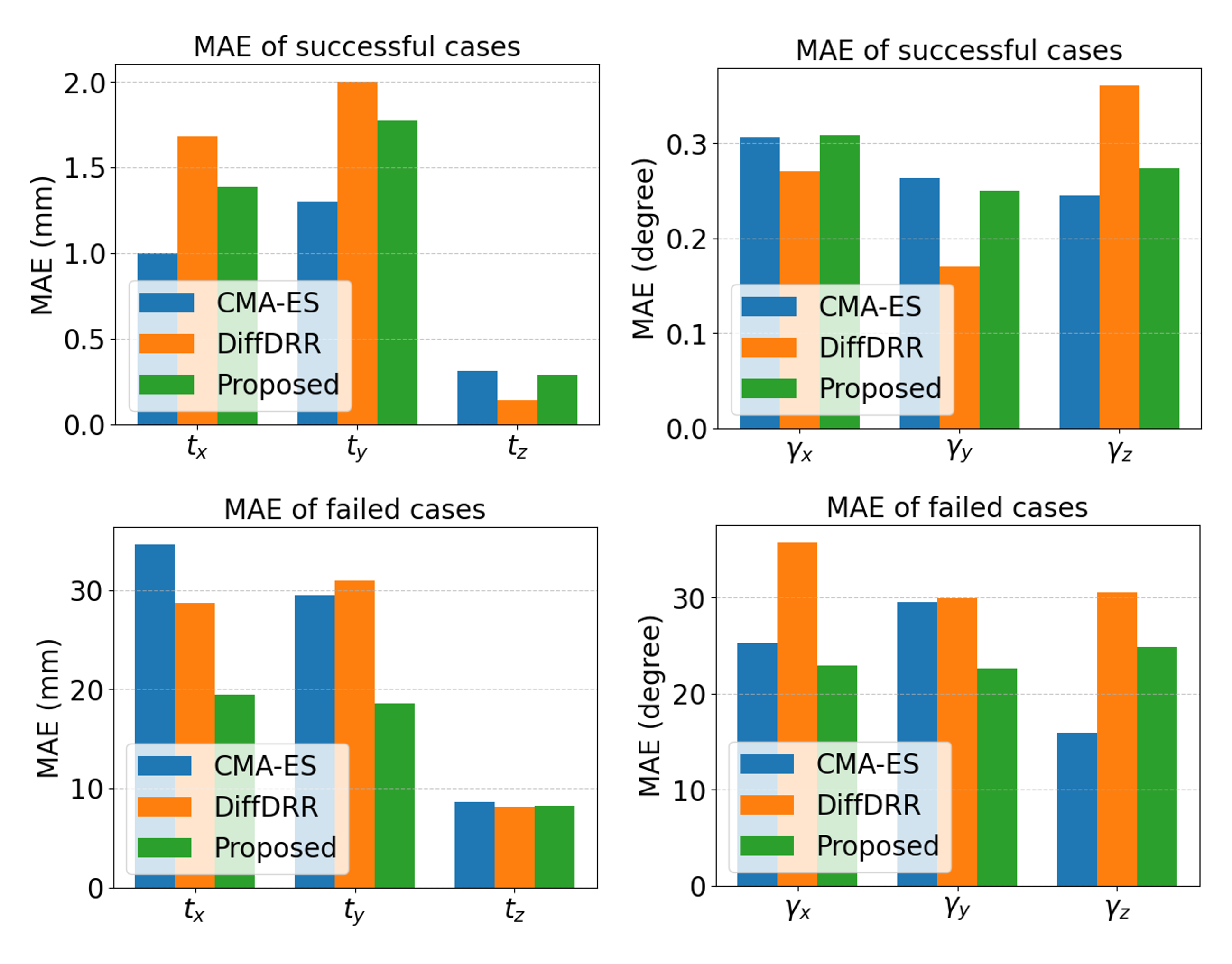}
    \caption{Mean absolute error (MAE) of the motion parameters estimated by CMA-ES, DiffDRR, and proposed approach. }
    \label{fig:2D3D}
\end{figure}

The 2D/3D registration results are summarized in Fig.\ref{fig:2D3D} and TABLE.\ref{tab:2D3D}. Gradient-based solutions (DiffDRR and Proposed) show comparable success rates, while CMA-ES shows $\sim8\%$ higher success. This improvement can be attributed to the fact that CMA-ES uses a population of solutions and is generally more robust in avoiding local minima and extending the capture range. In terms of computational efficiency, CMA-ES is $3.43\times$ slower than the proposed approach, and DiffDRR—also a gradient-based solution—is $7.80\times$ slower. For motion estimation accuracy in successful cases, the proposed approach achieves errors of approximately $\sim1.5mm$ for $t_x, t_y$ and $<0.5mm$ for $t_z$, which are comparable to CMA-ES and DiffDRR. Additionally, the proposed approach yields $\sim0.3\degree$ error in rotations estimation, also comparable to CMA-ES and DiffDRR. For failed registrations, the proposed approach achieves minimal errors on $t_x,t_y,\gamma_x,\gamma_y$. These results demonstrate the proposed approach can achieve 2D/3D registration performance comparable to existing gradient-free and gradient-based solution, while offering a significantly lower computational cost.

\begin{table}[h!]
    \centering
    \caption{2D/3D registration results using different optimization algorithms}
    \label{tab:2D3D}
    \renewcommand{\arraystretch}{1.5}
    \begin{tabular}{c|ccc}
        \Xhline{1.5pt}
        \textbf{Results}        & CMA-ES & DiffDRR & Proposed \\
        \hline
        \textbf{Runtime (mins)} & 63.39  & 143.95  & 18.45\\
        \hline
        \textbf{\# Success}     & 382    & 335     & 343\\
        \hline
        \textbf{\# Fail}        & 118    & 165     & 157\\
        \Xhline{1.5pt}
    \end{tabular}
\end{table}

\subsubsection{Motion-Compensated Analytical Reconstruction}
\begin{figure}
    \centering
    \includegraphics[width=1.02\linewidth]{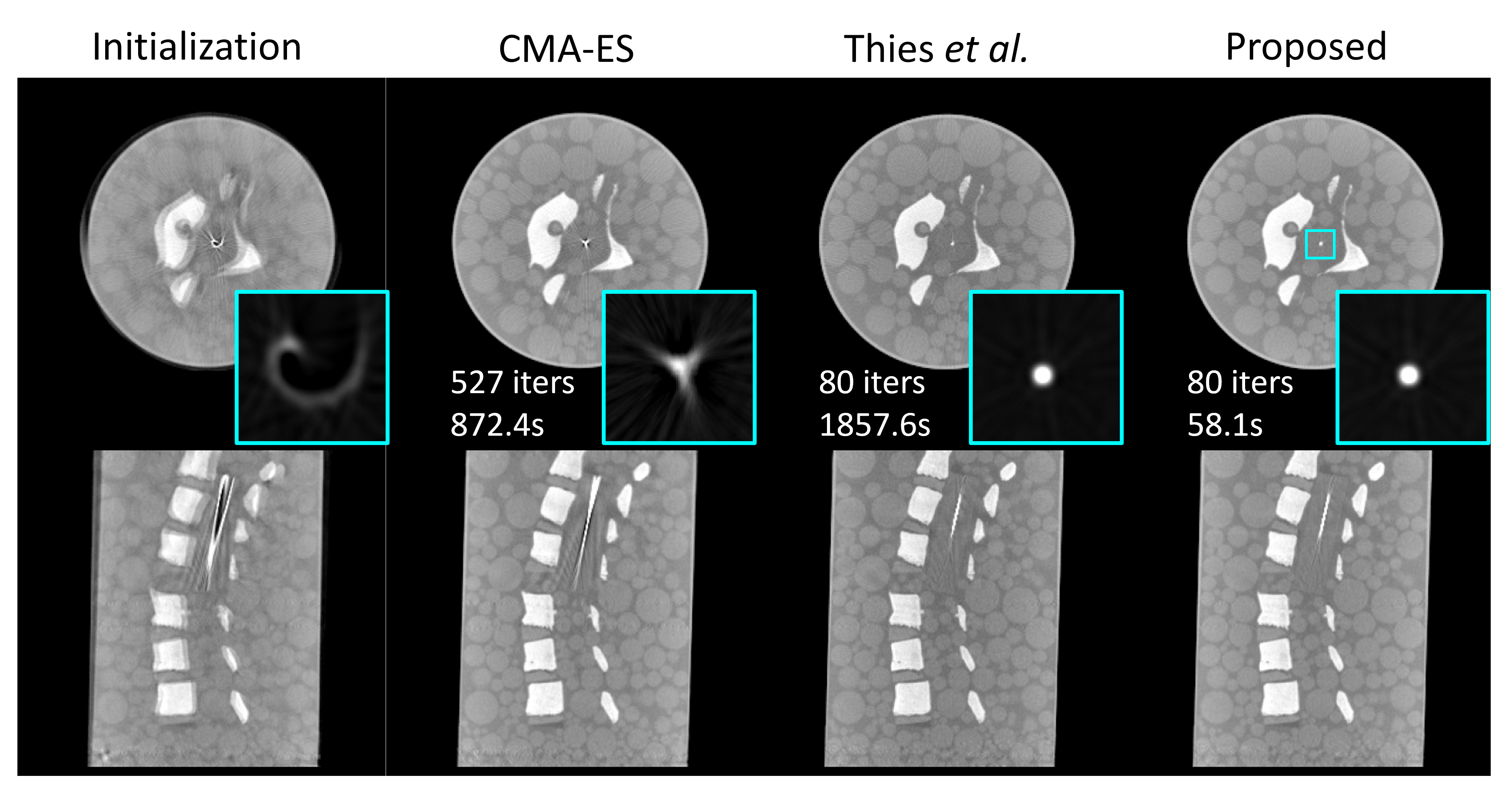}
    \caption{Motion compensated FBP reconstruction of a spine phantom. The full size volume is reconstructed with a standard resolution (0.5mm$^3$), and the zoom-in image containing a 0.127mm diameter tungsten wire, is reconstructed using the estimated motion with a high resolution (0.0625mm$^3$). Display window: full size/zoom-in: [0.01, 0.03]mm$^{-1}$/[0.015,0.1]mm$^{-1}$}
    \label{fig:MoCo_FBP}
\end{figure}
Fig.\ref{fig:MoCo_FBP} shows the motion-compensated FBP reconstruction results. The initial FBP reconstruction exhibits significant motion artifacts, such as blurred soft tissue, doubling of the high-density bone structures, and distortion of the wire, caused by phantom drifting and rotation. CMA-ES effectively estimates the motion parameters, restoring sharp edges of bony structures. The gradient-based methods (Thies \textit{et al} and Proposed) further enhance the motion compensation performance, with the wire structure clearly visualized in the coronal view. Using the estimated motion parameters, an ROI reconstruction with 8$\times$ resolution is performed, focusing solely on the wire region. These zoom-in images of the gradient-based solutions illustrate the auto-focusing of the central wire cross section, which finally achieves an isotropic reconstruction for the thin wire. Although the proposed approach produces reconstruction sharpness comparable to the method of Thies \textit{et al}, it achieves this with substantially shorter runtime. These results demonstrate the effectiveness of the proposed differentiable backprojector for motion-compensated analytical reconstruction in physical experimental data.

\subsubsection{Online Calibration of Noncircular Orbits}
\begin{figure*}
    \centering
    \includegraphics[width=\linewidth]{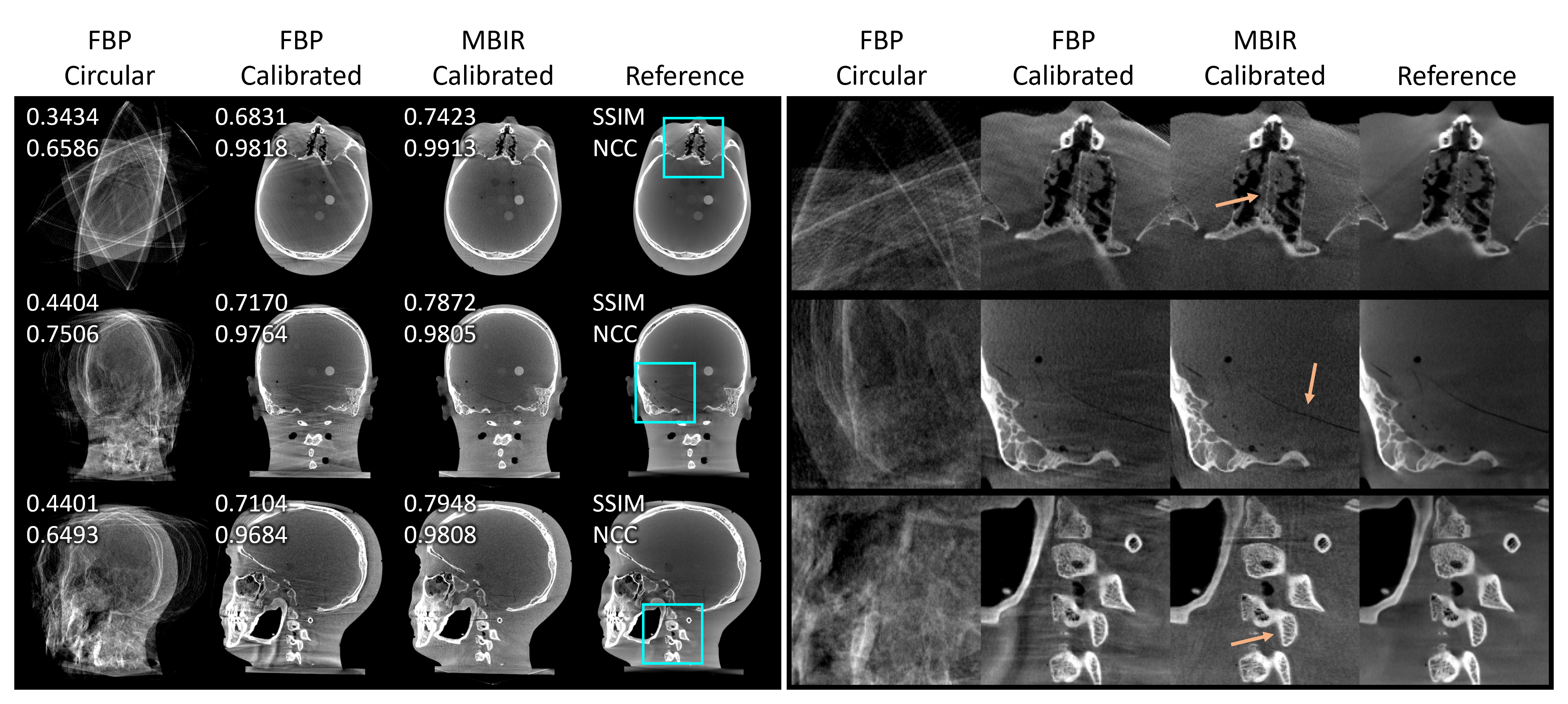}
    \caption{Reconstruction results using circular (initial) and calibrated geometries. The reference image was reconstructed using FBP from 1440 projections acquired with a circular scan. Structural Similarity Index (SSIM) and Normalized Cross-Correlation (NCC) are used to quatify the similarty between the reconstruction and reference image. The right panel shows zoomed-in views of bone ROIs to highlight the reconstruction quality of fine structural details and sharp features.}
    \label{fig:MoCo_MBIR}
\end{figure*}
Fig.\ref{fig:MoCo_MBIR} shows reconstruction results using the initial circular geometry and the online calibrated geometry. With the initial geometry, the reconstructed images exhibit severe degradation, and meaningful anatomical structures are barely discernible, indicating that the actual scan trajectory significantly deviates from the ideal circular orbit. In contrast, the online calibrated geometry leads to a substantial improvement in FBP reconstruction quality, as evident in both visual appearance and quantitative metrics. The MBIR method further suppresses streak artifacts in soft tissue, achieving the highest SSIM/NCC values across all views. The right panel provides zoomed-in views of bone ROIs to assess fine structural detail and image sharpness. Using the calibrated geometry, fine anatomical features indicated by arrows, such as cortical bone boundaries, trabecular textures, and soft-tissue crevices, are clearly resolved in both FBP and MBIR reconstructions. These results demonstrate the effectiveness of the proposed differentiable projector for accurate geometry calibration and high-quality reconstruction from real experimental data.

\section{Discussion and Conclusion}
In this work, we proposed a general and efficient framework for differentiable forward and back-projector with respect to rigid motion in X-ray imaging. By deriving an analytical expression for the motion gradient in the continuous domain, our approach circumvents the tedious formulation of path length and voxel footprint in the discretized domain. This continuous formulation reveals an important insight: the motion derivatives of forward and back-projector share the same integration structure as the original forward and back-projection operations, which allows existing projection algorithms to be directly adapted for computing motion gradients without relying on full auto-differentiation or being restricted to specific projector models. To further optimize performance in the discrete implementation, we introduce acceleration strategies based on weight sharing and path-based buffering, which effectively balances computational speed and memory usage. Evaluations demonstrate the numerical accuracy of the proposed method under various motion conditions, its computational advantages over existing differentiable projectors, scalability to large image sizes, and generalizability across different projector models. Validation on representative X-ray imaging tasks illustrates that integrating the proposed differentiable projectors into existing imaging frameworks enables effective gradient-based solutions, demonstrating their potential to improve a wide range of clinical applications.

The primary contribution of this work lies in the derivation of analytical projector gradients in the continuous domain, which forms the foundation for a concise mathematical formulation and a simple, adaptable implementation. Rather than computing the exact gradient of the discretized forward projector—an approach often used in conventional auto-differentiation-based frameworks—the proposed method instead derives the gradient analytically from a continuous perspective and subsequently discretizes the result for implementation. While this introduces a potential source of approximation error due to the mismatch between the continuous formulation and the discrete application, our empirical evaluations demonstrate that the resulting gradients align closely with those obtained via finite-difference methods. This observation validates the effectiveness of the proposed continuous-to-discrete strategy and the validation on multiple imaging tasks suggests that the proposed gradients are sufficiently accurate for optimization in a wide range of applications.

A key advantage of our approach lies in its memory efficiency. Traditional auto-differentiation techniques typically require the storage of extensive intermediate variables during the forward pass in order to compute backward gradients, leading to high memory consumption and limiting scalability. In contrast, our continuous-domain formulation eliminates the dependency on intermediate states by directly computing gradients based on analytical expressions, thereby reducing memory requirements. This enables our implementation to support high-resolution and large-scale scenarios that are otherwise infeasible using memory-intensive methods. For instance, our algorithm can successfully perform 2D/3D image registration tasks from a volumetric input of size 512$^3$ to projection images of size 768$^2$ using less than 3GB of GPU memory. This capability facilitates broader applications of differentiable projectors in computational imaging, including high-resolution multi-view registration\cite{liao2019multiview, ouadah2016self} and joint motion estimation and image reconstruction~\cite{de2025adaptive}. 

We investigated the performance of both gradient‑free and gradient‑based algorithms in 2D/3D registration and motion-compensated reconstruction. The results revealed that existing differentiable forward and back‑projector, as implemented in prior work, do not offer clear efficiency advantages over the well‑established gradient‑free CMA‑ES algorithm. In practice, CMA‑ES remains a strong baseline due to its robustness to local minima and ability to optimize the objective without requiring time-consuming gradient computation. Existing differentiable projectors, while enabling gradient‑based optimization, often suffer from large memory footprints and high runtime overhead when implemented via auto-differentiation or lack specific efficiency optimization, limiting their scalability to large‑scale 3D/2D registration or high‑resolution reconstruction. Our proposed differentiable projector framework addresses these limitations by deriving analytical motion gradients in the continuous domain and implementing them in a memory‑efficient, GPU‑optimized form. This allows gradient‑based methods to retain the accuracy of prior differentiable approaches while achieving superior computational efficiency over the gradient-free methods.

We evaluated the proposed approach by directly integrating the differentiable projector into several existing computational frameworks. Overall, the results were encouraging and demonstrated the general effectiveness of the proposed method. However, in certain cases such as the 2D/3D registration task, the optimization process occasionally failed to reach satisfactory solutions. These performance limitations are primarily attributed to the conventional hand-crafted loss functions, which tend to have narrow capture ranges and limited robustness against local minima. To mitigate this issue, the differentiable projectors could be incorporated into coarse-to-fine pyramid strategies \cite{thevenaz1998pyramid} or combined with population-based methods such as CMA-ES to enlarge the capture range. Moreover, due to its differentiable ability and superior memory efficiency of the proposed projectors, the proposed projector can be seamlessly combined with advanced deep learning pipelines \cite{huang2022reference, huang2024deformable, huang2023multi, gao2023fully}. Within such pipelines, the learned loss functions or data-driven motion priors can potentially address the capture range and convergence issues observed with hand-crafted losses. In this study, we simply used manually designed loss functions to validate the effectiveness of the proposed differentiable projector. Future work will explore integrating the proposed projector with learning-based approaches to fully exploit its potential in various X-ray imaging tasks. Additionally, we have recently investigated joint estimation problem in X-ray imaging solved by combining physical system modeling with advanced generative models\cite{lorenzon2025joint,jiang2025joint}. We note that a similar idea has been applied to motion estimation\cite{de2025adaptive,de2025solving}, but has been largely limited to sparse-view or low-resolution settings due to the extensive computational cost. We anticipate that the proposed differentiable projector could help alleviate the computational burden associated with these tasks.

The current study is limited to rigid object motion, however, the gradient formation in Eqs.~\eqref{eq:for_grad} and \eqref{eq:back_derive} are applicable to any motion model, provided a bilateral mapping between $\bs{r}$ and $\bs{r}'$ is defined, therefore, the proposed methodology can be directly extended to deformable motion, either via local rigid motion modeling\cite{capostagno2021deformable} or a fully deformable vector field modeling\cite{dang2015deformation}. While deformable motion requires more complex formulations and may significantly increase memory usage, future work will focus on adapting the framework to accommodate such scenarios. In particular, we aim to investigate parameterized deformable motion models to improve computational efficiency. Another promising direction is the extension of our framework to support advanced online geometry calibration. In this study, we employ a 6-DoF model, which assumes a fixed relative position between the source and detector. However, real-world geometric distortions may involve up to 9 DoF~\cite{ouadah2016self}, incorporating scaling and shearing components that cannot be represented as rigid object motion. Future work will explore a differentiable 9-DoF calibration approach to enable more accurate and flexible system modeling.

\bibliography{report} 
\bibliographystyle{IEEEref}

\end{document}